Letter

# Low temperature magnetic structure of CeRhIn$_5$ by neutron diffraction on absorption-optimized samples

D M Fobes[1], E D Bauer[1], J D Thompson[1], A Sazonov[2,3], V Hutanu[2,3], S. Zhang[4], F Ronning[1], and M Janoschek[1]

[1]MPA-CMMS, Los Alamos National Laboratory, Los Alamos, New Mexico 87545, USA
[2]Institut für Kristallographie RWTH Aachen University, 52056 Aachen, Germany
[3]Jülich Centre for Neutron Science (JCNS) at Heinz Maier-Leibnitz Zentrum (MLZ), Forschungszentrum Jülich GmbH, 85747 Garching, Germany
[4]Department of Physics and Astronomy, The University of Tennessee, Knoxville, Tennessee 37996, USA

E-mail: mjanoschek@lanl.gov

**Abstract.** Two aspects of the ambient pressure magnetic structure of heavy fermion material CeRhIn$_5$ have remained under some debate since its discovery: whether the structure is indeed an incommensurate helix or a spin density wave, and what is the precise magnitude of the ordered magnetic moment. By using a single crystal sample optimized for hot neutrons to minimize neutron absorption by Rh and In, here we report an ordered moment of $m = 0.54(2)\mu_B$. In addition, by using spherical neutron polarimetry measurements on a similar single crystal sample, we have confirmed the helical nature of the magnetic structure, and identified a single chiral domain.

## 1. Introduction

The Doniach phase diagram explores a competition in energy scales common to heavy fermion materials, between the Ruderman–Kittel–Kasuya–Yosida (RKKY) and Kondo interactions, which scale as $\sim J^2$ and $\sim e^{-1/J}$, respectively, where $J$ is the dimensionless Kondo coupling constant [1]. On one side of the phase diagram exists localized magnetism due to RKKY and on the other a totally Kondo-quenched paramagnetic state. Perhaps the most interesting scenarios are those in the middle of the phase diagram where the Kondo and RKKY energy scales are comparable. To find systems of this nature it is imperative to accurately measure these energy scales. CeRhIn$_5$ is one such system where these scales are believed to be competing.

The temperature–pressure phase diagram of CeRhIn$_5$ is prototypical for heavy fermion materials; at ambient pressure, it displays incommensurate antiferromagnetic (AFM) order below $T_N$ = 3.8 K, and with increasing applied pressure $T_N$ is suppressed to a quantum critical point at $P_c$ = 2.25 GPa around which a broad dome of unconventional superconductivity emerges [2-5]. In detail, however, the properties of the AFM ground state in CeRhIn$_5$ deviate from prototypical behavior. Namely, the ground state magnetic order was reported to be helical, rather than spin-density-wave-type (SDW) by neutron measurements [2] and was later confirmed by $^{115}$In NQR measurements [6, 7]. Notably, the Ce magnetic moments were found to lie in the tetragonal *ab*-plane with the helix propagating along *c* [8]. Recent neutron spectroscopy measurements observed a small energy gap $\Delta = 0.25$ meV in the spin wave spectra of the AFM state [9], but in the absence of magnetic anisotropies in the plane containing the spiraling spins a gapless Goldstone mode is expected for an incommensurate helical magnetic structure. Although an in-plane interaction invariant under the C$_4$ symmetry of the tetragonal structure may be possible, the gapless Goldstone mode is protected by translational symmetry of the magnetic helix along *c*. In principle, a sufficiently strong C$_4$ magnetic anisotropy (comparable to the exchange interactions along *c*), would distort the magnetic helix and break the associated translational symmetry, resulting in the formation of a spin gap. However, distortion of the helix would also lead to higher harmonic diffraction peaks, which have never been observed. Due to the size of the gap, the associated intensity of the peaks resulting from the distortion should, however, make them straightforward to observe. This is also in agreement with a calculation of possible exchange anisotropy terms based on the derived exchange Hamiltonian and the crystal field ground state that show that such anisotropies are too small to explain the size of the spin gap [10]. Similarly, the presence of crystallographic defects that break the translation invariance of the helix were theoretically considered to explain the gap, but also prove insufficient [9]. Therefore, a detailed reinvestigation of the previously determined magnetic structure is warranted.

In addition, while those same neutron spectroscopy measurements determined the RKKY energy scale in $CeRhIn_5$, conclusive information about the strength of the Kondo interaction has been difficult to ascertain. For example, an accurate measurement of the ordered magnetic moment may provide pertinent information about the strength of the Kondo interaction. Notably, a significant Kondo interaction may also provide an alternative explanation for the observed spin gap that does not rely on the existence of magnetic anisotropy. For temperatures smaller than the Kondo temperature it is well-known from neutron spectroscopy experiments on intermediate valence compounds [11-13] that the Kondo interaction gaps out low energy spin fluctuations. For this scenario, the size of the observed gap is directly determined by the Kondo temperature $T_K$. Although state of the art electronic structure methods such as dynamical mean field theory (DMFT) are currently not able to compute the effect of the Kondo interaction on the dynamic magnetic susceptibility in the magnetically ordered state, current estimates for $T_K$ based on bulk measurements, NMR and crystal field spectroscopy range from 5-28 K [14-17], suggesting that a Kondo gap of the observed size should exist. For a total absence of Kondo screening in $CeRhIn_5$, the ordered magnetic moment is expected to be 0.92 $\mu_B$, by calculation from crystalline electric field (CEF) excitations [14]. However, for $CeRhIn_5$ a wide range of values between 0.26 $\mu_B$ and 0.8 $\mu_B$ have been reported (c.f. Table 1) [6, 18-23]. Because these values were obtained below the temperature at which the order parameter saturates, it seems clear that the differences in these values are not from differences in temperatures, but rather from differences in analysis; because Rh and In are significantly neutron absorbing, special care, particularly at higher incident wavelengths, must be taken to adjust integrated intensities for absorption, which may account for the range of reported values.

In this report, we present data on these two debated aspects of the magnetic ground state of $CeRhIn_5$ obtained using hot neutrons. As we show in detail below, the use of hot neutrons overcomes analysis issues resulting from high neutron absorption and other related effects. Our results confirm the helical magnetic ground state of $CeRhIn_5$ and indicate a significant suppression of the magnetic moment, indicative of a Kondo interaction that is comparable to the RKKY exchange interactions even far away from the magnetic quantum critical point.

## 2. Experimental details

To overcome the issues related to the high absorption cross-section of both Rh and In, we opted to perform our neutron diffraction measurements utilizing hot neutrons, i.e. those exhibiting lower neutron wavelength. As seen in table 1, a lower wavelength results in a higher 1/$e$ absorption length, thus allowing for larger samples and better neutron statistics. For this reason, the neutron diffraction measurements were performed on the polarized neutron diffractometer POLI [24], situated at the hot neutron source of the FRM II reactor at the Heinz Maier-Leibnitz Zentrum. We selected a neutron wavelength of $\lambda = 0.7$ Å using a focused Si(311) monochromator, which is at the peak neutron flux of the hot source, and results in a 1/$e$ absorption length of ~4 mm, nearly double that of previous measurements. Additionally, with a lower neutron wavelength and the lifting counter available at POLI we can reach a larger region of reciprocal space, and neutron extinction is significantly lower; it has been previously shown that the extinction factor $y = I_{obs}/I_{kin}$, where $I_{obs}$ is the observed integrated intensity and $I_{kin}$ is the kinematic intensity, is roughly proportional to $1 - \lambda^2$ [25]. Therefore, the use of hot neutrons significantly improves the neutron statistics for the refinement of the magnetic structure of $CeRhIn_5$, and simultaneously decreases the amount of absorption and extinction corrections that must be applied during refinement. A single crystal of tetragonal $CeRhIn_5$ (space group P4/*mmm*) grown by the Indium self-flux method, with room temperature lattice parameters $a = 4.655$ Å and $c = 7.542$ Å [26], sample dimensions of 3.5 mm × 4.1 mm × 21.5 mm (~1 g), and mounted on a thin (0.5 mm) Al-plate with a hydrogen free glue (CYTOP), was used for the unpolarized experiment. A data set of 149 nuclear and 53 magnetic Bragg reflections, corresponding to 80 unique nuclear and 18 unique magnetic reflections, was collected for magnetic structure refinement at $T = 1.5$ K. The experimental structure factors of the measured Bragg reflections were obtained with the DAVINCI program [27]. Refinement was performed using single crystal refinement implemented in FullProf software [28]. Zero-field spherical neutron polarimetry (SNP) was performed using a third generation CRYOPAD [29], without a lifting counter, at $T = 0.4$ K. Polarized $^3$He spin filters are used for producing and analyzing neutron polarization. A similar single crystal of CeRhIn$_5$ with dimensions 3 mm × 3 mm × 2 mm (~150 mg) mounted on a Cu-post was used for the polarized experiment. Both crystals were mounted with an orientation such that the [HH0] and [00L] directions are in the scattering plane.

## 3. Results

To first confirm the helical nature of the magnetic structure we performed zero-field SNP measurements [30] on a small single crystal, performing an elastic scan along the (00$l$)-direction across a set of two magnetic Bragg peaks at $(-1/2, -1/2, -1 \pm \delta)$, where $\delta = 0.297$ describes the incommensurability of the helix along the $c$-axis. Considering the associated magnetic ordering vector $\boldsymbol{k} = (1/2, 1/2, \delta)$ and the constraint that the moments lie in the $ab$-plane [8], it can immediately be determined via representational analysis that the magnetization should take the form

$$\mathbf{M} = M\mathrm{Re}\big[(\mathbf{u} + \alpha \mathbf{v})e^{i2\pi n\delta}\big], \quad (1)$$

where $M$ is the magnitude of the magnetic moment, $\mathbf{u}$ and $\mathbf{v}$ are unit vectors perpendicular to $c$ and each other, $\alpha$ is a complex number, and $n$ is an integer [6]. Symmetry allows $\alpha = 0$ or $\pm i\beta$, where the former would result in a collinear spin-density-type modulation and the latter in a spiral structure. In the latter case, $\beta = 1$ represents a pure helix, whereas for $\beta < 1$ or $\beta > 1$ the helix exhibits an elliptical distortion. Although previous magnetic structure analysis by neutron diffraction selected the spiral structure using a Goodness-of-Fit parameter [6], the presence of a magnetic spiral can only be unambiguously determined using polarized neutrons [31-35]. SNP measurements allows the determination of the polarization tensor that describes how the polarization of the incident neutron beam changes during the scattering process due to the dipole-dipole interaction between the neutron spin and the magnetic moments in the investigated sample, as defined by the Blume-Maleyev equations [31-35]. Here the incident polarization can change in three distinct ways; (i) it can rotate around the Fourier transform of the magnetization of the sample, (ii) its magnitude can be reduced due to depolarization of the beam from magnetic domains, or (iii) the sample can act as a neutron polarizer and generate a new component of the neutron polarization. A magnetic helix can be easily distinguished from a spin-density wave via SNP measurements because it generates polarization of the neutron beam parallel to the direction of the scattering vector $\boldsymbol{Q}$ due to a finite chiral contribution to the magnetic scattering cross-section [36]. In SNP, the coordinate frame is typically defined to rotate with $\boldsymbol{Q}$ so that $\boldsymbol{x}$ is parallel to $\boldsymbol{Q}$, $\boldsymbol{z}$ is perpendicular to the scattering plane and $\boldsymbol{y}$ completes a right-hand set. Thus, by measuring any element of the polarization tensor with the final polarization being along $\boldsymbol{x}$, we can unambiguously confirm or eliminate the possibility of a spiral structure. Notably, any non-zero contribution in the spin flip channel of the polarized scattering cross section $\sigma_{xx}$ can unambiguously confirm the presence of the chiral contribution due to a spiral spin structure. However, because a magnetic helix may be left- or right-handed, in principle two chiral domains may exist. If the chiral domains with opposite handedness exhibit the same volume ratio the total measured chiral contribution will be zero. Therefore, the presence of the chiral scattering can also serve as an indicator for the chiral domain population.

In Fig. 1 we present the two spin-flip cross sections $\sigma_{xx}^{\uparrow\downarrow}$ and $\sigma_{xx}^{\downarrow\uparrow}$ for the set of peaks $(-1/2, -1/2, -1 \pm \delta)$ corrected for the time-dependent decay of the polarized spin filters, and subsequently subtracted by a similarly corrected background. Determination of and correction for the time-dependence of the $^3$He spin filters was performed as established in Ref. [37]. Diffraction scans were performed at 0.4 K, well below the order parameter saturation temperature of ~2 K. In both panels of Fig. 1, intensity in only one scattering cross section was observed for each peak, i.e. $\sigma_{xx}^{\uparrow\downarrow}$ and $\sigma_{xx}^{\downarrow\uparrow}$ for $(-1/2, -1/2, -0.703)$ and $(-1/2, -1/2, -1.297)$, respectively, demonstrating that the chiral contribution to the magnetic cross-section for CeRhIn$_5$ is non-zero. This clearly demonstrates the presence of a helicoidal magnetic structure, as opposed to a sine-modulated structure, which was allowed by symmetry, and may have potentially explained a previously observed spin-wave gap [9]. We can estimate the chiral domain population via the chiral ratio $r_{chir}$ using the expression [38]

$$r_{chir} = \frac{\sigma_{xx}^{\uparrow\downarrow} - \sigma_{xx}^{\downarrow\uparrow}}{\sigma_{xx}^{\uparrow\downarrow} + \sigma_{xx}^{\downarrow\uparrow}},$$

where the cross sections $\sigma_{xx}^{\uparrow\downarrow}$ and $\sigma_{xx}^{\downarrow\uparrow}$ are shown in Fig. 1, for each of the two peaks. Using the corrected peak counts and adjusting to the calculated filter efficiencies, we obtain chiral ratios of $r_{chir} = 0.99(4)$ and $0.96(6)$, for Fig. 1a and 1b, respectively, demonstrating that the neutron beam is fully polarized along the $x$ direction after scattering, thus indicates that the utilized sample effectively consists of a single chiral domain, and thus is homo-chiral. We note that this behavior is unique, in that strictly speaking one might expect equal chiral domain populations in an unstrained centrosymmetric crystal, due to both domains being energetically equivalent. Although the origin of this observation is unclear, it is consistent with a previous report which reported unequally populated chiral domains [23].

The previously determined nuclear structure, with space group P4/*mmm* [26], was utilized to refine the crystallographic parameters on the nuclear Bragg peaks. An absorption correction ($I = I_0 e^{-\mu d}$, where d is the diameter of the sample) was performed prior to refinement using a linear absorption coefficient of $\mu = 0.104 \text{ mm}^{-1}$, and an extinction correction was performed during refinement using the Becker-Coppens method, as implemented in FullProf [39], where the extinction parameter $y$ was nearly unity for the majority of refined peaks. The resultant structural parameters are presented in Table 2. The low temperature values of $a = 4.606$ Å and $c = 7.467$ Å, obtained at $T = 0.4$ K, both differ from the previously reported room-temperature values by ~1%, representing a reasonable thermal relaxation of the lattice. The results of the nuclear refinement are shown in Fig. 2a, as the calculated versus observed squared structure factors of the measured nuclear Bragg peaks.

The magnetic moment was refined using the previously determined helicoidal structure, as we confirmed above by SNP [6], for which the cross section is given by

$$\sigma(\boldsymbol{q}) = \frac{1}{4}\left(\frac{\gamma r_0}{2}\right)^2 \langle M \rangle^2 |f(Q)|^2 (1 + |\hat{\boldsymbol{q}} \cdot \hat{\boldsymbol{c}}|^2), \qquad (2)$$

where $(\gamma r_0/2)^2 = 0.07265$ barns/$\mu_B^2$, $M$ is the staggered moment, $f(Q)$ is the magnetic form factor, and $\hat{\boldsymbol{c}}$ is the unit vector along the $c$-axis. Since we have confirmed the spiral spin structure, the magnetization takes the form of Eq. 1, where $\alpha = \pm i\beta$. Using the results from the structure refinement of the nuclear Bragg peaks, we obtain an ordered moment of $\langle M \rangle = 0.54(2)\mu_B$, with Goodness-of-Fit parameter $\chi^2 = 5.46$, compared to $\chi^2 = 4.19$ for the nuclear structure. Fits with $\alpha = \pm i$, i.e. $\beta = 1$, result in the best Goodness-of-Fit parameters. The result of the magnetic refinement is similarly shown in Fig. 2b. The obtained value $\langle M \rangle = 0.54(2)\mu_B$ lies well within the range of the previous results (cf. table 1) [6, 18-23]. However, the results reported in Refs. [6, 18] not only suffered from the absorption and extinction problems described above, but also used an incorrect pre-factor in Eq. 2, later corrected in Ref. [22]. This narrows the range of previously reported moments to 0.34-0.8 $\mu_B$ [19-23] (cf. table 1). Two points are noteworthy; (i) the measurements reported in Ref. [19, 21] were carried out in pressure cells, which typically reduces the signal-to-noise ratio, making an accurate determination of the size of magnetic moments more difficult, and (ii) measurements performed with longer wavelengths seem to report larger magnetic moments. Extinction effects, which decrease the intensity of strong nuclear peaks and thus overestimate the magnetic moment, explain the latter point; as described above, extinction effects are significantly worse for longer wavelengths. There is one notable exception to this trend; the lowest moment of 0.34(5) $\mu_B$ was measured with the longest wavelength of $\lambda = 2.385$ Å [19], 45% larger than in any of the other experiments. Because neutron absorption increases exponentially with $\lambda$, this becomes more important than extinction (which increases with $\lambda^2$) in the limit of larger wavelength, resulting in the significant reduction of weak Bragg peaks, such as magnetic peaks, and thus an underestimated magnetic moment. Therefore, the disagreement for previously obtained values for the magnetic moment in CeRhIn$_5$ are easily explained, and we can similarly demonstrate that our result, obtained with an experimental setup and sample designed to overcome both absorption and extinction issues, is reliable.

In Fig. 3 we plot the quantity $\sigma(\mathbf{q})/(1 + |\hat{\mathbf{q}} \cdot \hat{\mathbf{c}}|^2)$, which should be proportional to the square of the Ce$^{3+}$ magnetic form factor, as seen in Eq. 2 [40]. We also show the Rh$^+$ magnetic form factor [41], illustrating consistency with previous reports that the magnetic moments reside on the Ce ions rather than Rh. The deviation from the reported pure Ce$^{3+}$ form factor at low $Q$ may be attributed to hybridization of the localized Ce 4$f$ electrons with the conduction electrons via Kondo coupling [42]. We note that in previous measurements [6] this enhancement of the form factor was not observed, likely because the experiment was carried out with more than two times larger neutron wavelength for which absorption corrections become more important.

## 4. Discussion

Our polarized neutron diffraction results identify a non-zero chiral contribution to the magnetic scattering and thus demonstrate unambiguously that the low-temperature magnetic structure of CeRhIn$_5$ is helimagnetic. Moreover, the measured chiral ratio shows the used sample of CeRhIn$_5$ is homo-chiral. We note that this also in agreement with a recent polarized neutron diffraction on CeRhIn$_5$ reported in Ref. [23]. Although the confirmation of the spiral magnetic state anticipates a gapless spin wave spectrum, our unpolarized hot magnetic neutron diffraction data also provide a valid scenario to explain the presence of the observed spin gap [9]. Notably, the strongly reduced magnetic moment and the deviation from the Ce$^{3+}$ magnetic form factor both suggest a Kondo interaction that is substantial fraction of the RKKY interaction. Of course, transvers spin fluctuations may also reduce the magnitude of the ordered moment. However, based on the spin-wave calculation used to fit our recent spin wave measurements in the ground state of CeRhIn$_5$ [9], we find that due to spin fluctuations the ordered moment should only be reduced by 17% compared to the full moment. In contrast, our current results demonstrate that the ordered magnetic moment is reduced by 41% suggesting significant Kondo screening. In principle, in the absence of magnetic anisotropies in the plane in which the magnetic moments of the helix spiral, the spin wave associated with an incommensurate spiral magnetic ground state should be a gapless Goldstone mode. However, the significant Kondo screening in CeRhIn$_5$ identified here suggests the existence of strong longitudinal magnetic fluctuations. Similar to insulating quantum magnets in which longitudinal fluctuations are induced by spin dimerization [43, 44] this longitudinal mode is expected to be critically damped and corresponds to a massive (gapped) Higgs mode. The spin waves observed in Ref. [9] are indeed damped. We note that the energy scale that determines the size of the spin gap associated with this longitudinal mode arises from the competition between Kondo and RKKY interactions. Further, the spin gap must vanish at the magnetic quantum critical point where the staggered magnetic moment goes to zero is approached via the application of external pressure. Consequently, careful measurements of the spin wave gap as function of pressure are required [10]. This is also consistent with the above Kondo scenario because the strength of Kondo coupling is expected to decrease with increasing $H$. Finally, current estimates for the Kondo temperature $T_\mathrm{K}$ in CeRhIn$_5$ are in a range from 5-28 K [14-16], which would be sufficient to verify the scenario proposed here.

## 5. Conclusion

Using hot neutrons, thus avoiding the complications of neutron absorption and extinction, we have established the ordered moment of heavy-fermion material CeRhIn$_5$ $\langle M \rangle = 0.54(2)\mu_B$ and confirmed that the magnetic ground state is helicoidal via spherical neutron polarimetry. These results provide a timely and necessary refinement of a series of contested results, and will aid in the theoretical understanding of this material. In particular, the 41% reduction of the magnetic moment compared to 0.92 $\mu_B$ expected from the crystal field ground state, together with the enhancement of the Ce magnetic form factor at large $Q$ established by our measurements, suggest that even in the magnetically ordered state the magnetic moments are significantly Kondo screened in CeRhIn$_5$. Similarly, sizable Kondo screening is also suggested by NMR measurements (5 K ~ 0.4 meV) [17], and by the large critical magnetic field $H \sim 50$ T (~5 meV) required to saturate the AFM state with a low $T_N = 3.8$ K (~0.3 meV) [45]. Current theory treats the magnetically ordered state in heavy fermion materials in such a way that the Kondo interaction is negligible compared to the exchange interaction, but our results suggest that new theories which effectively incorporate a sizable Kondo interaction, even in the magnetically ordered state, are urgently needed to make progress on the understanding of the spin Hamiltonian and the emergent properties of heavy fermion materials.

 Acknowledgements
We gratefully acknowledge useful discussions with Cristian Batista, Shizeng Lin and Igor Zaliznyak. Work at Los Alamos National Laboratory (LANL) was performed under the auspices of the U. S. Department of Energy. Research supported by the U.S. Department of Energy, Office of Basic Energy Sciences, Division of Materials Sciences and Engineering under the project "Complex Electronic Materials", apart from DMF who was funded by the LANL Directed Research and Development program. LANL is operated by Los Alamos National Security for the National Nuclear Security Administration of DOE under contract DE-AC52-06NA25396.

**Table 1**. Previous reported values of the ordered magnetic moment with the utilized neutron wavelengths and corresponding $1/e$ absorption length. We note that the measurements in Refs. [12, 14, 15] were performed in a pressure cell. Ordered moments from Refs. [6, 18] were calculated using an incorrect prefactor, later corrected in Ref. [22].

| Magnetic Moment ($\mu_B$) | Neutron Wavelength (Å) | $1/e$ absorption length (mm) | Ref. |
|---|---|---|---|
| 0.26 | 1.53 | 1.7 | [6] |
| 0.37 | 1.53 | 1.7 | [18] |
| 0.8 | 1.53 | 1.7 | [19] |
| 0.59 | 1.28 | 2.1 | [20] |
| 0.6 | 1.64 | 1.6 | [21] |
| 0.75 | 1.53 | 1.7 | [22] |
| 0.34 | 2.39 | 0.95 | [23] |

**Table 2**. Structural parameters at $T$=1.5 K. (Note: $\chi^2 = 4.19$).

| | |
|---|---|
| $a$ | 4.606(1) Å |
| $c$ | 7.467(2) Å |
| $z$ | 0.3055(1) Å |
| $u_{Ce}$ | 0.0094(4) Å$^2$ |
| $u_{Rh}$ | 0.0099(3) Å$^2$ |
| $u_{In1}$ | 0.0114(3) Å$^2$ |

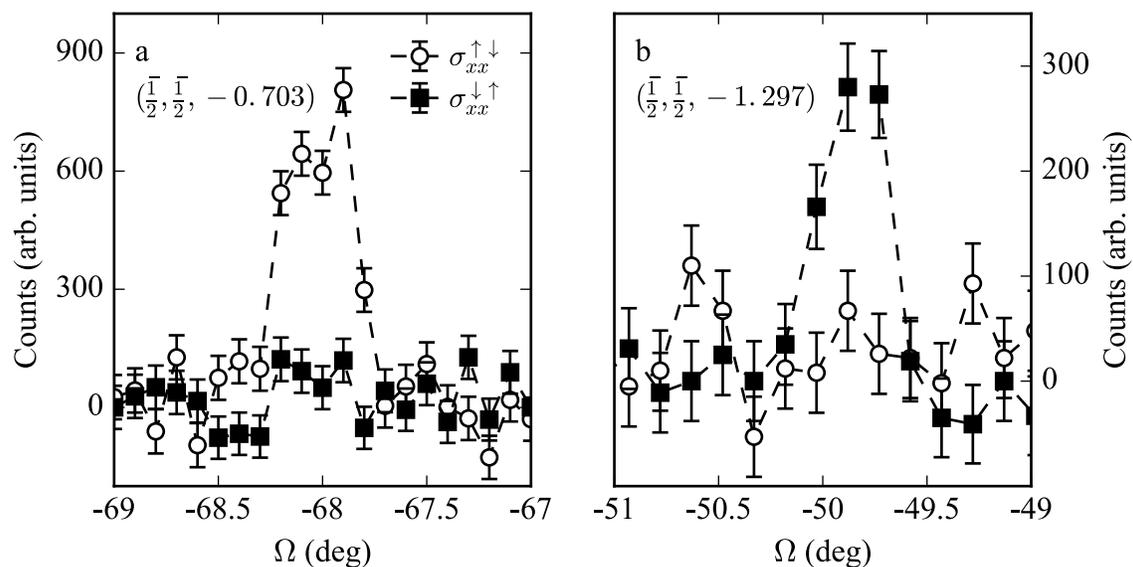

**Figure 1**. Normalized neutron counts obtained via spherical neutron polarimetry with CRYOPAD of the *xx* component of the polarization tensor at a temperature $T = 0.4$ K, where circles are $\sigma_{xx}^{\uparrow\downarrow}$ and squares are $\sigma_{xx}^{\downarrow\uparrow}$ as a function of the rocking angle $\Omega$ for the a) (0.5, 0.5, 0.703) and b) (0.5, 0.5, 1.297) magnetic Bragg peaks. Raw counts were adjusted for the decay of the $^3$He spin filters and background subtracted.

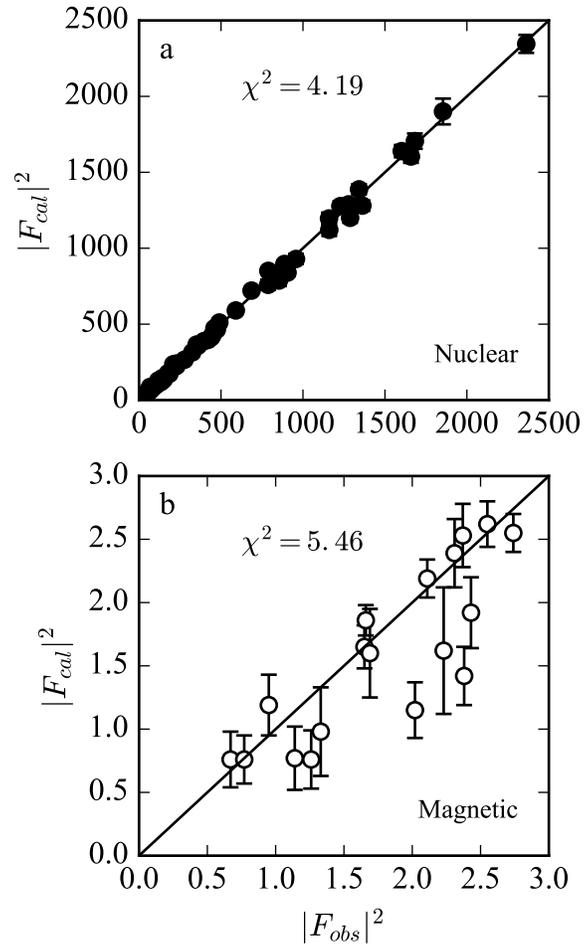

**Figure 2**. Results of single crystal refinement of a) the nuclear peaks and b) magnetic peaks, displayed as the observed vs. calculated squared structure factors.

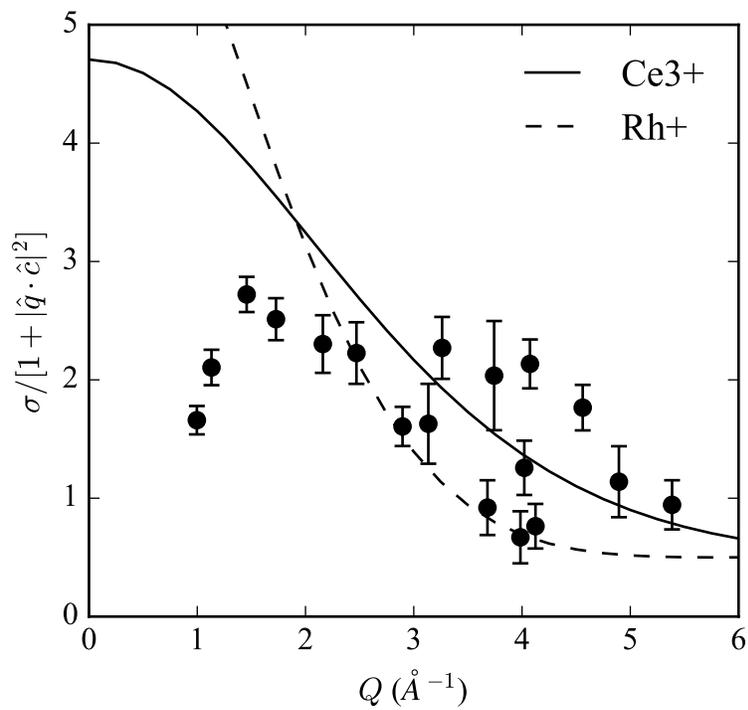

Figure 3. The *Q*-dependence of the magnetic cross section obtained at a temperature *T* = 0.4 K divided by the polarization factor. The solid line is the $Ce^{3+}$ magnetic form factor [40], and the dashed line the $Rh^{+}$ magnetic form factor [41].